# Numerical modelling of the effect of weathering on the progressive failure of underground limestone mines

Siavash GHABEZLOO[1], Ahmad POUYA
*Laboratoire Centrale des Ponts et Chaussées, Paris, France*

ABSTRACT: The observations show that the collapse of underground limestone mines results from a progressive failure due to gradual weathering of the rockmass. The following stages can be considered for the limestone weathering and degradation process in underground mines: condensation of the water on the roof of the gallery, infiltration of the water in the porous rock, migration of the air $CO_2$ molecules in the rock pore water by convection and molecular diffusion, dissolution of limestone by $CO_2$ rich water and consequently, reduction of the strength properties of rock. Considering this process, a set of equations governing different hydro-chemo-mechanical aspects of the weathering phenomenon and progressive failure occurring in these mines is presented. Then the feasibility of numerical modelling of this process is studied and a simple example of application is presented.

## 1 INTRODUCTION

Many underground limestone mines exist in France particularly in the Parisian and Normandy regions. The failure of these underground mines causes damages on infrastructures and buildings on the ground surface and sometimes their total ruin. An overview of the existing methods for modelling the failure of underground mines shows that they cannot predict adequately the progress rate and the occurrence time of this phenomenon. The objective of this paper is first to present a set of equations governing different hydro-chemo-mechanical aspects of the progressive failure process occurring in these mines. Then, the feasibility of numerical modelling of this process will be studied and a simple example of application will be presented.

## 2 FAILURE MECHANISM

Observations on various cases show a typical type of failure that emerges brutally to the ground surface by creating a pseudo-circular crater (Fig 1 d), called "fontis", begins with a localized failure at the roof of the gallery (Fig 1a). The rise of the bell shaped rupture to the roof of the gallery generally follows the local ruptures of the roof. This process can evolve to stability due to presence of a resistant and stiff layer that is opposed to the development of failure. The rise of the rupture thus stops as long as there are not any more new degradations on the level of the high-roof. With time, the resistance of the stable layer could be degraded and the failure process continues until the rupture emerges on the surface, in the form of a crater (Tritsch et al. 2002, Abbas Fayad 2004). The diameter and the depth of the "fontis" can vary between a few meters and several tens of meters according to the geometry of the mine, nature and the thickness of the covering layers and the presence of a water table in these layers.

The results of in-situ experiments of Morat et al. (1999) can help to understand the mechanism of gradual degradation and failure in the underground limestone mines. They observed that if mirrors are placed near the floor, walls (or pillars) and roof of a cavity, it is apparent that water condenses on the roof and the upper part of the pillars but not at the bottom of the pillars nor on the floor where mirrors remain dry. Furthermore, deposits of calcite on the floor and lower part of the pillars constitute indirect evidence of water evaporation. They carried out the experiments in the Mériel quarry, 46 Km northwest of Paris, dug according to the 'rooms and pillars' method. The porosity of limestone was in the 25 - 45% range (Jouniaux et al. 1996) and the water saturation was between 60 and 80%. Temperature, air pressure and relative humidity measurements in the gallery and self-potential recordings in the walls of gallery evidenced the oscillatory convective motions of the air of the gallery, driven by the geothermal gradient, which transports water and heat. The temperature difference between the floor and the roof of the gallery is lower than that which would give the normal geothermal gradient, indicating that the transport of heat is more effective in the air of the gallery than through the wall-rock. Consequently, they noted that the water, which evaporates on the floor of the gallery, is transported by convection to the roof where it condenses and enters the rock. This water attacks the rock, preferentially along the existing cracks. The water can invade these cracks, enlarge them by dissolution and cause the failure of the roof. The process will repeat itself and in some cases can cause the appearance of the fontis on the ground surface. The mechanism of the gradual failure is summarized in the figure (1). The time needed for the completion of this process depends on the thickness of the competent layer. For limestone quarries in the Paris region, it seems to vary from a few tens of years to a few centuries (Morat et al. 1999).

---
[1] Presently in CERMES, Ecole Nationale des Ponts et Chaussées, Paris, France
    Email: ghabezloo@cermes.enpc.fr

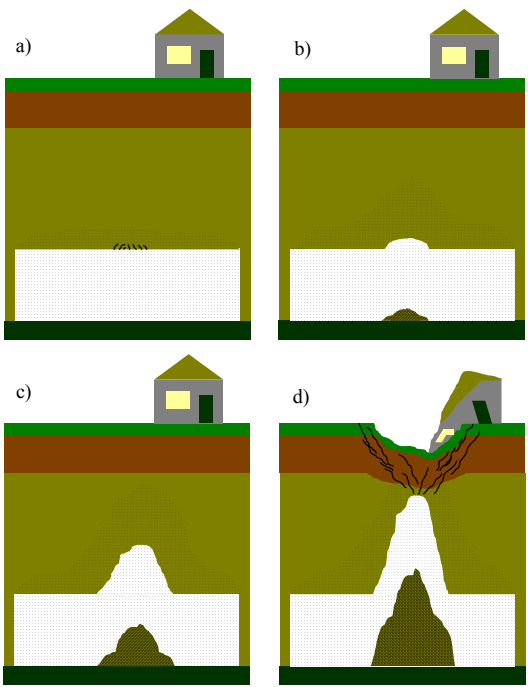

Figure 1. Mechanism of gradual failure of underground limestone mines.

## 3 PHYSICO-MECHANICAL PROPERTIES OF LIMESTONE

Massieu (1984), Thierry (1987) and Pothérat et al. (2002) studied the physico-mechanical properties of the limestone of the underground quarry of Villiers-Adam, located in Val d'Oise. Measurements of degree of saturation show that the samples taken in the roof have the highest degrees of saturation, between 85 and 100%. The degree of saturation of the pillars ranges between 58 and 72% and the separated blocks on the floor have the smallest degree of the saturation, located between 49 and 57%. These results are coherent with the evaporation and condensation process, described by Morat et al. (1999). The petrographic analysis shows that the limestone of the underground quarry of Villiers-Adam is constituted approximately of 90% calcite. More than four hundred porosity measurements show that the values of porosity are between 28 and 46% with an average value of 37,7%. The results of the uniaxiale compression tests show the strong influence of the degree of saturation on the compressive strength of limestone (figure 2). The results of the tests on the samples with different porosities show the decrease of the compressive strength of the saturated rock with porosity increase (figure 3). An increase in the porosity from 30 to 40% involves a reduction in the compressive strength from 13 to 4.5 MPa. The results of Brazilian tests show the same influence of the porosity on the tensile strength of saturated limestone (figure 3). $R_c$ and $R_t$ in the figures (2) and (3) are respectively uniaxiale compressive strength and tensile strength. The values measured for the Young's modulus were relatively dispersed, varying between 4600 and 12000 MPa. The results shown in figure (4) indicate that this modulus decreases for higher porosities.

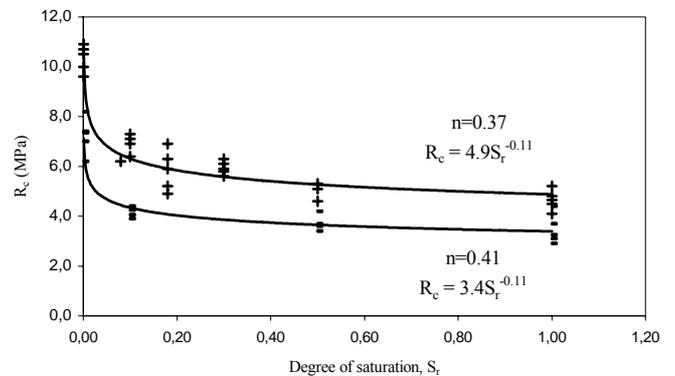

Figure 2. Reduction of compressive strength by degree of saturation (Thierry 1987)

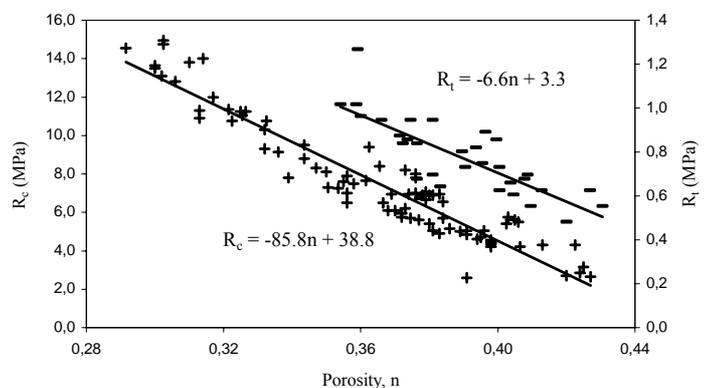

Figure 3. Reduction of compressive and tensile strength by porosity for saturated samples (Thierry 1987)

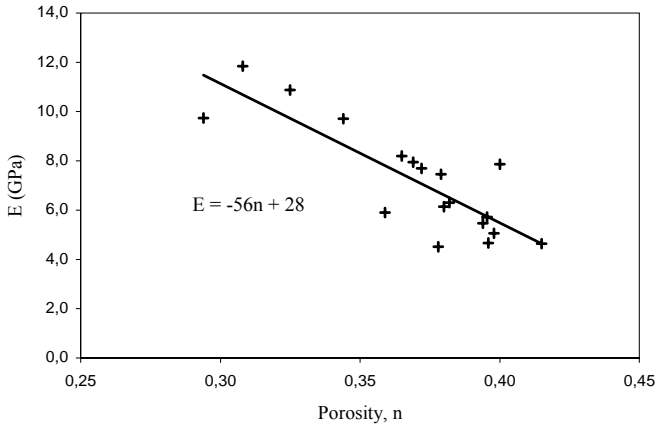

Figure 4. Reduction of Young's modulus by porosity for saturated samples (Thierry 1987)

## 4 PROPOSED METHOD: EFFECT OF WEATHERING

As described before, the observations show that the collapse of underground limestone mines results from a gradual phenomenon of weathering of the rockmass leading to a progressive failure.

### 4.1 *Weathering process*

Considering the in-situ experiments of Morat et al. (1991) in the limestone quarry of Mériel and the studies of Massieu (1984), Thierry (1987) and Pothérat et al. (2002) on the physico-mechanical properties of the limestone of the underground quarry of Villiers-Adam, we can consider the following mechanism for the weathering of limestone in underground cavities. Hajna (2002) observed a more or less identical mechanism in several natural cavities of limestone and dolomite in Slovenia.

Water evaporates on the floor and condenses on the roof of the cavity. Consequently, water is always present on the surface of the roof and infiltrates in the unsaturated rock. Infiltration of water in the rock causes the increase of the degree of saturation and as shown before, it decreases the strength properties of the rock. The pure water condensed on the roof dissolves the air's $CO_2$ and becomes acidic. The quantity of $CO_2$ dissolved in unit volume of water depends on its partial pressure in the cavity air and the temperature. So the water infiltrating the rock is rich in $CO_2$ and acidic and causes dissolution of calcite. Besides, the $CO_2$ concentration on the surface is higher than within the rockmass. This concentration gradient causes an inward migration of $CO_2$ from the rock surface by molecular diffusion. Consequently, the concentration of saturation of $Ca^{2+}$ ions in the rock pore water increases and the limestone continues to dissolve in its pore water. By this mechanism of limestone dissolution, the porosity increases and consequently the mechanical degradation continues even without more water infiltration. To include all these mechanisms in the model, we consider four stages in the weathering process of limestone: 1) in-filtration of water in the rock which increases the degree of saturation, 2) inward migration of dissolved $CO_2$ from the roof surface by advection and molecular diffusion, 3) dissolution of limestone, 4) mechanical degradation or reduction in the strength parameters of limestone due to the increase in the degree of saturation and also in the porosity of the rock.

By introducing the kinetic model and characteristic times of these different processes in the global model, we can elaborate a model for the weathering process of limestone and gradual failure of the underground cavities including the temporal aspect. Thus, it is necessary to study four different phenomena and try to model them: 1) water transfer, 2) diffusion-advection, 3) limestone dissolution, 4) mechanical degradation. In the following sections we describe separately these phenomena and their models.

### 4.2 *Water transfer phenomenon*

The hydraulic head is deduced from the potential of water and is written:

$$H = \frac{S}{\gamma_w} + z = h + z \qquad (1)$$

Where $H$ is the hydraulic head, $S$ is the suction, $\gamma_w$ the specific weight of water, $z$ is the elevation and $h$ is the pressure head. To calculate the suction according to the water content, the following equation is used (Kaufmann 2003):

$$S_r = e^{\alpha h} \Rightarrow h = \frac{1}{\alpha} \ln\left(\frac{\theta}{n}\right) \qquad (2)$$

Here $\theta$ is volumic water content, $S_r$ is degree of saturation, $n$ is porosity and $\alpha$ is an experimental constant.

$$\theta = S_r n \qquad (3)$$

Darcy's law is assumed to remain valid in unsaturated medium and the permeability is supposed to be constant:

$$V = -K \nabla H \qquad (4)$$

Where $V$ is velocity and $K$ is permeability. The continuity equation expresses the mass conservation for the water in the case of incompressible unsaturated medium:

$$\nabla \cdot V + \frac{\partial \theta}{\partial t} = 0 \qquad (5)$$

## 4.3 Diffusion-advection phenomenon

The transport of the solids dissolved in rock pore water can occur via two mechanisms, advection and diffusion. The advection results from the movement of water while diffusion occurs because of gradients of concentration in the medium (Fetter 1999). The total mass of the dissolved materials per unit of area, transported per unit of time is described by the following equation:

$$F = v\theta C - D\theta \nabla C \quad (6)$$

Where $v$ is mean velocity, $C$ is molarity (M/L$^3$) and $D$ is coefficient of diffusion (M$^2$/T). By writing the equation of mass conservation for a representative elementary volume and considering the mass produced by chemical reactions we find:

$$\frac{\partial(\theta C)}{\partial t} + \nabla \cdot F + R = 0 \quad (7)$$

Where $R$ is the reaction rate of calcite dissolution defined as the mass produced by chemical reactions in unit volume of water per unit of time.

## 4.4 Dissolution phenomenon

### 4.4.1 Dissolution mechanism

Limestone dissolves in water and the presence of carbon dioxide increases the calcite solubility. The dissolution of calcite in water in presence of carbon dioxide is given by the following reaction (Plummer et al. 1978, Plummer and Busenberg 1982, Dreybrodt and Gabrovsek 2003):

$$CaCO_3 + CO_2 + H_2O \rightarrow Ca^{2+} + 2HCO_3^- \quad (8)$$

In this reaction, for each calcite molecule released from the rock, a carbon dioxide molecule is consumed by conversion into bicarbonate ions. The result of the equilibrium chemistry is the equation describing the molarity of $Ca^{2+}$ in equilibrium according to the $CO_2$ partial pressure, $P_{CO2}$. (Dreybrodt and Gabrovsek 2003, Dreybrodt 1988, Kaufmann 2002):

$$\left[C_{Ca^{2+}}\right]_{eq} = \left(P_{CO_2} \frac{K_1 K_C K_H}{4 K_2 \tilde{\gamma}_{Ca^{2+}}^2 \tilde{\gamma}_{HCO_3^-}^2}\right)^{\frac{1}{3}} \quad (9)$$

In this equation $K_1$, $K_2$, $K_C$ and $K_H$ are empirical mass balance coefficients that depend on the temperature. $\tilde{\gamma}_{Ca}$ and $\tilde{\gamma}_{HCO3}$ are respectively the ion activity coefficients of calcium and bicarbonate (Plummer & Busenberg 1982, Kaufmann 2002). The $CO_2$ molarity, $C_{CO2}$ is given by:

$$C_{CO_2} = K_H P_{CO_2} \quad (10)$$

If the coefficients $\tilde{\gamma}_{Ca}$ and $\tilde{\gamma}_{HCO3}$ in equation (9) are taken equal to 1, then this equation and (10) allow us to write:

$$\left[C_{Ca^{2+}}\right]_{eq}^3 = \frac{K_1 K_C}{4 K_2}\left[C_{CO_2}\right] \quad (11)$$

### 4.4.2 Dissolution kinetics

The rate of dissolution of calcite depends mainly on chemical saturation of water and can be shown by this empirical equation (Kaufmann 2003, Palmer 1991 & 2003, Dreybrodt et al. 1996):

$$R = \frac{dC}{dt} = k_{cal}\left(1 - \frac{C_{Ca^{2+}}}{\left[C_{Ca^{2+}}\right]_{eq}}\right)^{n_{cal}} \quad (12)$$

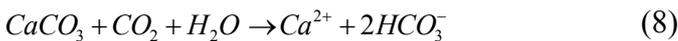

Where $k_{cal}$ is the reaction coefficient (M/L$^3$T) and $n_{cal}$ is the order of reaction. $C_{Ca2+}$ and $[C_{Ca2+}]_{eq}$ are respectively molarity and saturation molarity of dissolved calcite in water (M/L$^3$). The ratio of these terms is called saturation ratio. The values of $n_{cal}$ and $k_{cal}$ change with saturation ratio, temperature and $CO_2$ partial pressure.

## 4.5 Degradation phenomenon

The dissolution of calcite increases the volume of voids and thus the porosity of rock. The porosity increase is equal to the volume of dissolved calcite per unit of fluid volume, multiplied by the volumic water content:

$$dn = \frac{dV_v}{V_t} = \frac{dV_{cal}}{V_t} = \theta \frac{m}{M} dC_{Ca^{2+}} \quad (13)$$

In these relations, $dV_v$ is the variation of volume of the voids, $V_t$ is total volume and $dV_{cal}$ and $dC_{Ca2+}$ are respectively the variation of volume and the number of moles of dissolved calcite. $m$ is mass of a mole of limestone and $M$ is mass of one litre of limestone. Using the equations (12) and (13), we obtain:

$$\frac{dn}{dt} = \theta \frac{m}{M} \frac{dC_{Ca^{2+}}}{dt} = \theta \frac{m}{M} R \quad (14)$$

The mechanical degradation in our model has two different origins. The first one is the increase in degree of saturation. The second one is the porous matrix dissolution and porosity increase. As a matter of fact, in laboratory experiences if the water injected in a rock sample is not chemically saturated with the rock minerals, then it can also dissolve these minerals. In this case the two effects mentioned above contribute simultaneously to mechanical degradation of the rock. To separate the parts of strength reduc-tion due to these mechanisms, specific testes have to be realised with chemical control of the water injected to the sample or passing through and leaching it. In the data given in figures (2) and (3) the effects of the two mechanisms of mechanical degradation are superposed. Since we have not other data allowing us to separate the part of the two mechanisms, we admitted as a work assumption, that the reduction in the mechanical parameters can be attributed to the first effect (degree of saturation increase) for 80%, and to the other one (dissolution) for 20%. Supposing the partial pressure of carbon dioxide, $P_{CO2}$, equal to 0,003 atm (Palmer 1991) and using equation (9) we can deduce from the data of the figure (2) the reduction of the compressive strength versus the dissolved mass of limestone (Figure 5). We calculated from the data of the figure (2) for n=0.37, the quantity of dissolved calcite corresponding to the volume of water that enters the rockmass and increases the degree of saturation.

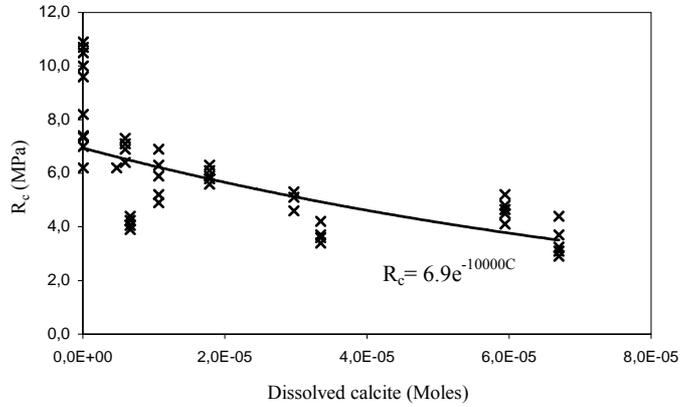

Figure 5. Variation of the compressive strength due to dissolution of calcite deduced from the data of figure (2)

Considering an exponential function for the strength reduction and using the assumption described above for attributing 20% of the strength reduction to dissolution phenomena, we can calculate the rate of variation of compressive strength as a function of the mass of dissolved calcite:

$$R_C = 6.9 e^{-0.2 \times 10000 C} \Rightarrow \frac{dR_C}{R_C} = -2000 \, dC \qquad (15)$$

According to the data of figure (3), the tensile strength is about 12% of the compressive strength; therefore we can also use the equation (15) to evaluate the tensile strength reduction.

## 5 APPLICATION TO THE CAVITY FAILURE

To study the effect of weathering on the gradual failure of the underground limestone mines we consider an underground cavity at 20m depth. The geometry of the model is presented in figure (6). For the numerical modelling we use the part delimited by the dashed lines (square of 9m).

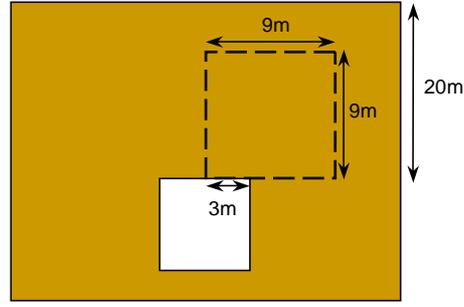

Figure 6. Geometry of model

### 5.1 Initial conditions

To define the initial conditions of the molarity of $Ca^{2+}$ in the limestone mass before digging of the cavity, we consider the dissolution in a closed system with the initial partial pressure of $CO_2$ equal to 0.003 atm (Palmer 1991). The initial value of molarity of $Ca^{2+}$ is calculated equal to $1.607 \times 10^{-4}$ mole/litre. The initial value of the molarity of $CO_2$ in the mass of limestone is null, but to avoid the numerical difficulties, we take a very low value. The initial porosity of limestone in the solid mass is constant. To define the hydraulic initial conditions, we calculate the values of the volumic water content such that the water transfer velocity in the equation (4) is null, in other words, there is no flow of water before the digging of the cavity.

### 5.2 Boundary conditions

As it was mentioned above, the in-situ measurements of Morat et al. (1999) show that water evaporates on the floor and condenses on the roof of the cavity. Consequently, water is always present in the roof surface. This corresponds to a boundary condition of degree of saturation equal to 1 on this surface, or the volumic water content is equal to the porosity.

On the roof of the cavity, carbon dioxide dissolves in pure water. The dissolved quantity of $CO_2$ per unit volume of water depends on its partial pressure in the atmosphere and on the temperature of the cavity. Thus we calculate the molarity of $CO_2$ in the boundary of the model by using equation (10). Considering the partial pressure of $CO_2$ equal to 0.003 atm and the temperature equal to 10°C, we obtain the $CO_2$ molarity equal to $1.611 \times 10^{-4}$ mole/litre.

Since the $Ca^{2+}$ ions cannot leave the roof, we impose the condition of null flow in the vertical direction on the lower limit of the model and thus the term $F$ given by the equation (6) is equal to zero and we obtain:

$$n_r \frac{\partial C_{Ca^{2+}}}{\partial r} + n_z \frac{\partial C_{Ca^{2+}}}{\partial z} = \frac{V_r C_{Ca^{2+}} n_r + V_z C_{Ca^{2+}} n_z}{\theta D} \qquad (16)$$

We impose the conditions of null gradient on the higher boundary and the boundaries on the left and on the right of the model, therefore the gradients of water content and $Ca^{2+}$ and $CO_2$ molarities are null also on these boundaries.

## 5.3 Final equations

For the numerical modelling, it is initially necessary to finalize the equations of the problem. Four phenomena were considered for the process of weathering of limestone: water transfer, diffusion-advection, dissolution and degradation. For the phenomenon of water transfer we use the equations of Darcy and the conservation of mass. Using the equations (1) to (5), we can calculate the flow velocity and the water content in all of the nodes. We assumed that the $Ca^{2+}$ and $HCO_3$ ions are always in equilibrium. That means, according to (8), the molarity of $HCO_3$ ions is always twice the molarity of $Ca^{2+}$ ions. With this assumption, we write the equations of diffusion-advection only for the ions $Ca^{2+}$ and the molecules of $CO_2$. Using the equations (5), (6), (7), (11) and (12), and considering the variables $C_1$ and $C_2$ respectively for the molarities of $CO_2$ and $Ca^{2+}$, we obtain:

$$\theta \frac{\partial C_1}{\partial t} + \nabla \cdot (V C_1) - \nabla \cdot (D_1 \theta \nabla C_1) - C_1 \nabla \cdot V + R = 0 \quad (17)$$

$$\theta \frac{\partial C_2}{\partial t} + \nabla \cdot (V C_2) - \nabla \cdot (D_2 \theta \nabla C_2) - C_2 \nabla \cdot V - R = 0 \quad (18)$$

These equations and initial and boundary conditions constitute a complete set of equations and conditions allowing us to do a numerical modelling.

## 5.4 Model parameters

The values of the model parameters are presented in Table (1). We have not found sufficient data in the literature to determine the value of the coefficient $k_{cal}$ in equation (12) for the dissolution process in the pore structure of limestone. The results are very sensible to this coefficient. For high values of this coefficient the dissolution rate is very high and the chemical weathering is concentrated in very vicinity of the roof surface. The value given here for this coefficient was chosen as to find a zone of active weathering process of about a few tens of centimeters depth from the roof surface.

Table 1. Model parameters

| Parameter | Value | Unit | Parameter | Value | Unit |
|---|---|---|---|---|---|
| $n_{ini}$ | 0.4 | - | $C_{2ini}$ | $1.607 \times 10^{-4}$ | mole/litre |
| $C_{1ini}$ | $3.982 \times 10^{-7}$ | mole/litre | $K$ | 0.005 | m/year |
| $C_{1lim}$ | $1.611 \times 10^{-4}$ | mole/litre | $D_1$ | 0.3 | m²/year |
| $dt$ | 0.002 | Year | $k_{cal}$ | $1.0 \times 10^{-3}$ | mole/litre-year |
| $D_2$ | 0.3 | m²/year | $\alpha$ | 0.03 | m$^{-1}$ |
| $n_{cal}$ | 4 | - | $m$ | 100 | g |
| $M$ | 2500 | g | $K_1$ | $10^{-6.46}$ | mole/litre |
| $K_C$ | $10^{-8.41}$ | mole²/litre² | $K_2$ | $10^{-10.49}$ | mole/litre |

## 5.5 Modelling results

A 2D program is developed using MATLAB software and the Finite Differences method. A grid containing 2025 nodes is created. The smallest distance between the nodes in the horizontal and vertical directions is respectively 10 and 5 cm near the roof of the cavity. By using the model equations we can calculate the rate of strength reduction with the time. To have a simple idea of the failure zone, a linear elastic mechanical modelling is done using CESAR-LCPC software and the values of tensile vertical stress are calculated in the nodes corresponding to the nodes of current model. By calculating the ratio of tensile stress to tensile strength in each node in the model we can have an idea of the failure zone. We ignore the stress field changes due to cracking and also the reduction of the elastic parameters with dissolution process. The first step of the modelling was stopped when the first clear zone of failure is appeared in the 80$^{th}$ year and the results are as follows:

Figure (7) shows the volumic water content. It shows an increase in the first few meters of depth of the rock, where the water condenses on the roof. Figures (8) and (9) show respectively the $CO_2$ and $Ca^{2+}$ molarities. The transport of the dissolved $CO_2$ molecules and produced $Ca^{2+}$ molecules in the rock with the advection and diffusion mechanisms is clearly shown.

Figure (10) shows the porosity increase in the model. As it can be seen, the maximum increase is in the order of 10$^{-4}$ that is a very low value. The more important phenomenon that we can see in this figure is the higher porosity increase in the corner of the roof, where we have the instantaneous changes in the boundary conditions. This is a numerical problem that we could not completely resolve and causes the appearance of the first failure zones in this zone, which is not compatible with the observations in underground mines.

Figure (11) shows the values of tensile strength. We can see the significant decrease in the tensile strength in the first meter of the roof depth, where it can be a possible zone for the failures.

Figure (12) shows the ratio of tensile vertical stress to tensile strength. The maximum value of the ratio is considered as 1 and we assume the failure is happen in these nodes. After 80 years, we could see the failure zone showed in the figure (12) with the maximum depth of about 20cm and we assume it as the first step of the gradual failure of cavity.

For the modelling of the second failure step, another mechanical modelling is done with the new geometry of the model and the vertical tensile stresses were calculated. By transferring the boundary conditions to the new boundaries the modelling was continued until the appearance of another failure zone. That happens in the 125th year with the failure zone showed in figure (13).

These results show that by using the presented equations it is possible to model limestone weathering and gradual failure of the underground mines with the progress rate and failure shape more or less compatible with the observations of the different cases of failure in these mines.

Obviously with a completely coupled finite element model, a more refined mesh and specially with the more precise parameters we can have a better model to predict the progressive failure of the underground cavities.

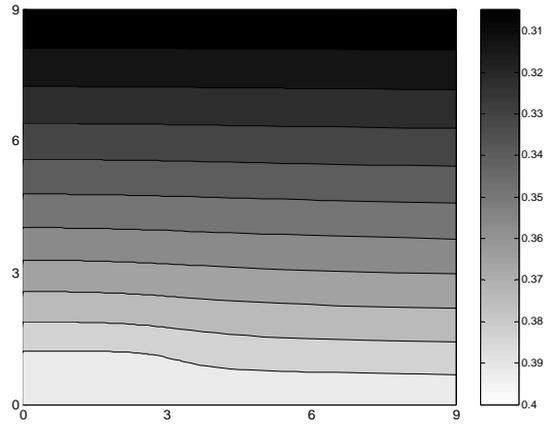
Figure 7. Volumic water content – 80th year

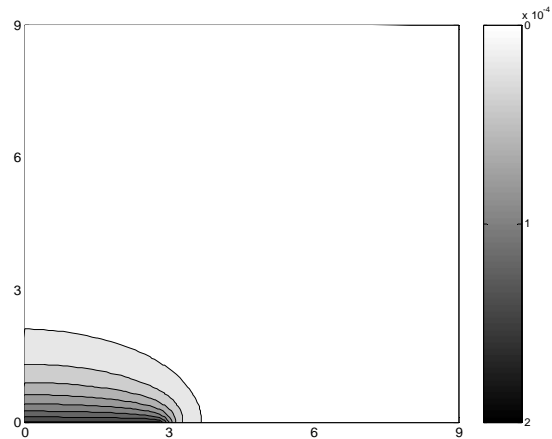
Figure 8. Molarity of $CO_2$ – 80th year

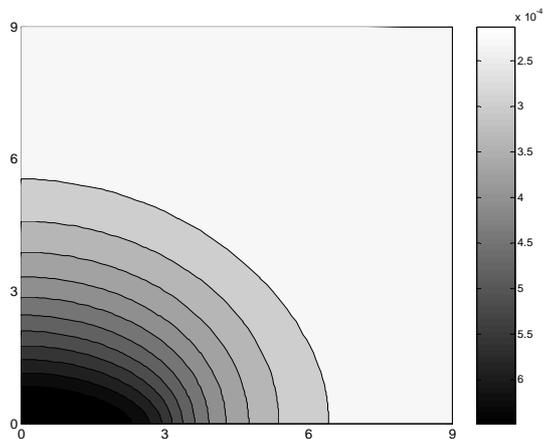
Figure 9. Molarity of $Ca^{2+}$ – 80th year

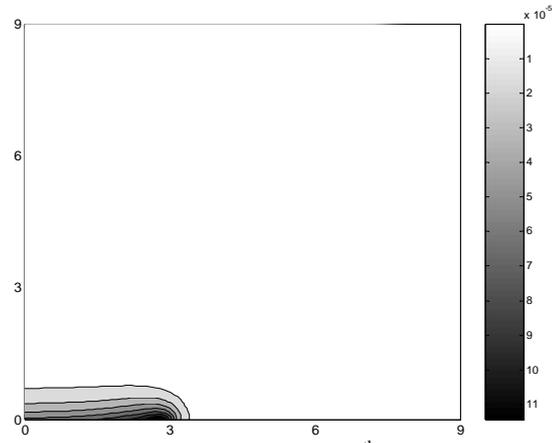
Figure 10. Porosity increase – 80th year

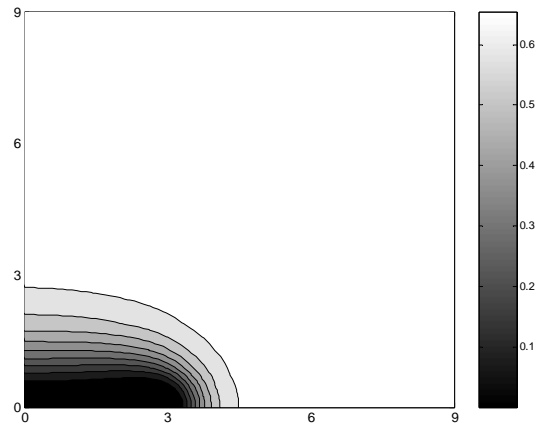
Figure 11. Tensile strength – 80th year

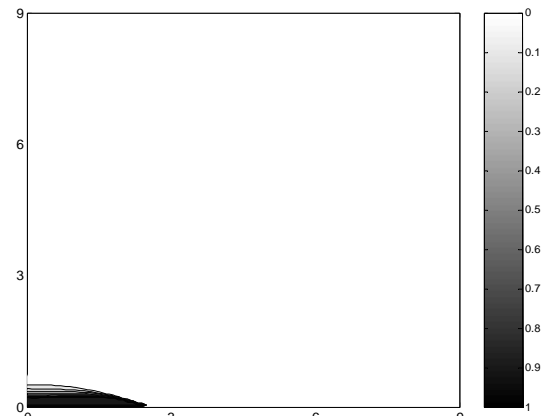
Figure 12. Tensile stress/tensile strength – 80th year

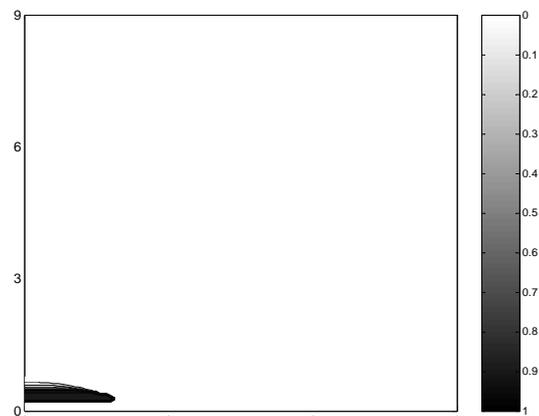
Figure 13. Tensile stress/tensile strength – 125th year

# 6 CONCLUSION

The progressive failure in underground mines is partly due to weathering processes that take place under natural conditions within and in the vicinity of the rockmass. The weathering process consists of chemo-physical mechanisms that have a mechanical degradation effect for the rock. We showed that by taking into account the kinetics of the weathering process, the transport of chemical constituents in the rockmass and the model of mechanical degradation caused by chemo-physical interactions, it is possible to give a model of progressive failure for underground structures. The work presented in this paper is based obviously on very sharp simplification assumptions. The aim of this work was only a first approach of this problem and to show the feasibility of such a modelling. This work must be continued by using more adequate data and models and also compared to *in situ* observations. The very interesting result of such an approach would be a kinetic model of the progressive failure and so an access to the temporal aspect of the failure in risk assessment of underground mines.